# Maxwell's Demon and Its Fallacies Demystified

Milivoje M. Kostic

*Professor Emeritus of Mechanical Engineering*
Northern Illinois University, USA
Email: kostic@niu.edu ; Web: www.kostic.niu.edu

A demonic being, introduced by Maxwell, to miraculously create thermal non-equilibrium and violate the Second law of thermodynamics, has been among the most intriguing and elusive wishful concepts for over 150 years. Maxwell and his followers focused on 'effortless gating' a molecule at a time, but overlooked simultaneous interference of other chaotic molecules, while the demon exorcists tried to justify impossible processes with misplaced 'compensations' by work of measurements and gate operation, and information storage and memory erasure with entropy generation. The illusive and persistent Maxwell's demon fallacies by its advocates, as well as its exorcists, are scrutinized and resolved here. Based on holistic, phenomenological reasoning, it is deduced that a Maxwell's demon operation, against natural forces and without due work effort to suppress interference of competing thermal particles while one is selectively gated, is not possible at any scale, since it would be against the physics of the chaotic thermal motion, the latter without consistent molecular directional preference for selective timing to be possible. Maxwell's demon would have miraculous useful effects, but also some catastrophic consequences.

## Introduction

"*Only simple qualitative arguments can reveal the underlying physic*" was quoted by Philippe Nozieres and 'heartily agreed' by Anthony Legget, a Nobel laureate [*Science Bulletin* 63 (2018) 1019-1022], emphasizing "mathematical convenience versus physical insight … that theorists are far too fond of fancy formalisms which are mathematically streamlined but whose connection with physics is at best at several removes." In light of that, with phenomenological insight and simple, holistic reasoning, the elusive Maxwell's Demon (MD) fallacies, as wishful concepts for over 150 years, are critically examined and demystified here. It is hoped that this treatise will also help demystify some recent challenges of the Second Law of thermodynamics and promote constructive future debates.

James Clerk Maxwell introduced a hypothetical 'intelligent being (very observant and neat-fingered micro-being)' in 1867 [1], and publicly in 1871 [2], to miraculously and effortlessly partition the chaotic thermal molecules of a gas in equilibrium into faster and slower groups, and thus, by effortlessly creating a non-equilibrium with work potential, challenge the Second law of thermodynamics (SL), see Figs. 1 & 2. William Thomson (Lord Kelvin) was the first to use the word '*demon*' for Maxwell's concept in 1874 [3]. Maxwell's demon has challenged many, including among the best creative and scientific minds in thermodynamics and statistical physics, chemistry and biology, quantum mechanics, information theory and cybernetics, and philosophy and sociology. Demon embodiment morphed into many weakly defined or fictional structures, has been kept alive with creative imagination, and defended by approximate and incomplete arguments, based on newly developing concepts, with inconsistent and incoherent evidences. Its utility has never been successful.

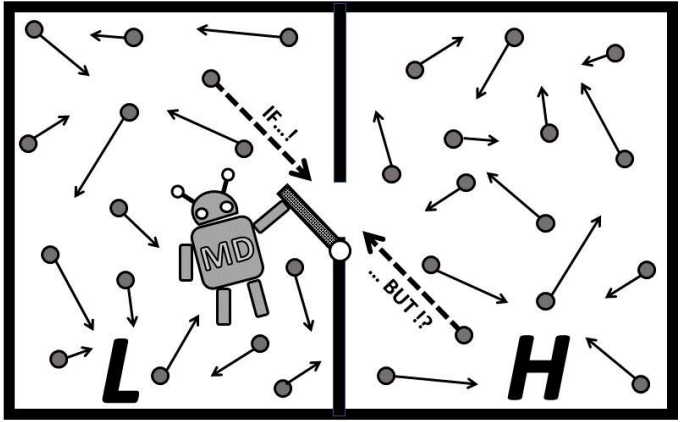

**Fig. 1: Maxwell's demon (MD) operates its gate.** MD opens the gate to 'wishfully pass' a higher speed molecule from *L* to H (see dashed arrow line in *L*) and lower speed molecules in reverse. However, considering the chaotic and fast molecular thermal motion (most molecules are faster than sound speed, see Fig. 2), it is probable that the same or even higher speed molecule from *H* will pass back to *L* in that time period or collide with an oncoming molecule (see dashed arrow line in *H*). Even higher speed molecules may pass back from *H* to *L*, and more probably so if MD was 'successful by chance' to separate more high-speed molecules into *H*. Therefore, 'just opening the gate' would 'more equalize than separate' by speed, due to simultaneous molecular interactions and interference, see Fig. 3.

The science and technology have evolved over time on many scales and levels, so that we now have advantage to look at its historical developments more comprehensively and objectively than the pioneers. Still, the human curiosity and ingenuity persistently result in wishful imagination (or unrealistic 'thought experiments') even if against the forces and processes of nature. Sometimes, highly accomplished scientists and authors in their fields, do not fully appreciate the essence of the subtle Second Law of thermodynamics.

Maxwell's demon has been a topic of many scientific, philosophical and social publications, some in the most prestigious scientific journals, even editorials [4]. In a dedicated, one volume anthology [5], Leff and Rex (2002) presented a wide-ranging overview of the Maxwell's demon's life and status, with reprints of important original papers, including an extensive and annotated bibliography with selected citations, with 570 entries in alphabetical and chronological lists. However, the authors [5] chose to emphasize the progress of the MD developments and missed an excellent opportunity to present all alternatives and to encourage due debate, particularly since most of the SL challenges have been resolved in favor of the SL and never against. Therefore, the goal here is not to review the MD literature, but to provide a phenomenological reasoning, independent from scale size and description approach.

Regardless of never-ending obstacles and controversies, but due to many creative and some mystical writings, mostly scattered throughout diverse literature, the MD is still alive, and is a motivation and inspiration for challenging the SL of thermodynamics. Fascination with the demon has accelerated throughout the development of statistical and quantum physics, and information theory. However, the vast majority of scientific community up to now have not been convinced with the "demonologists' (MD protagonists).

## Persisting Maxwell's Demon Fallacies

Four persisting fallacies of the MD advocates and exorcists are reasoned and presented here, namely: (*i*) Overlooking simultaneous interference of other molecules with Maxwell demon's selected molecule while passing through the gate (Figs. 1, 2 & 3); (*ii*) Ignoring physical impossibility of processes against the natural forces, e.g., impossibility of 'free compression' (Fig. 4); (*iii*) Inappropriate justification of SL validity by compensation elsewhere and later (e.g., let's justify destruction of entropy locally -- alluding spontaneous heat transfer from lower to higher temperature -- and compensate it by entropy generation elsewhere and/or later (Fig. 5); and (*iv*) Inappropriate justification of temperature reduction as cooling by heat transfer from lower to higher temperature (alluding to SL violation), although it is possible to be done without heat transfer, by adiabatic (ideally isentropic) expansion (Appendix, Fig. A1). Additionally, miraculous and wishful MD actions would result in infinite refrigeration or engine efficiencies, but also in unforeseen catastrophic consequences (Fig. 6), among others.

## Self-tendency towards Thermal-Equilibrium, Interference Due-Work and Operational Gate-Work

Stable and perpetual thermal macro-equilibrium consists of simultaneously-and-randomly interacting (i.e., forcibly interfering) molecules (thermal particles) with minimum free-energy and maximum entropy for an established equilibrium state, resulting in random distribution of position and momenta of molecules, well-approximated by the Maxwell-Boltzmann distribution over a wide range of speeds, with extremely high and low values (curve *T* 's on Fig. 2). The stable equilibrium is self-sustained and resilient for any change, and an external work would be required for re-establishing a nonequilibrium, the latter will possess work potential equal to the reversible work, i.e., the most efficient, minimum work needed to create it, that could ideally be reversed back.

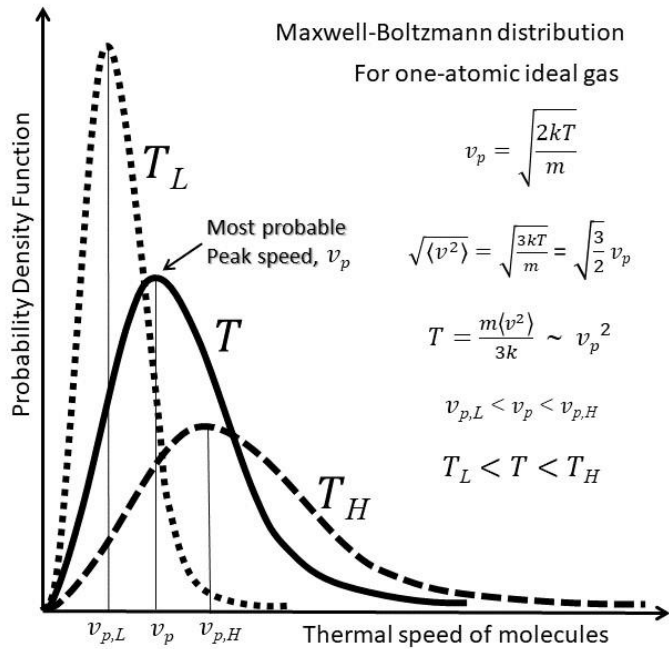

**Fig. 2: Self-sustained, Maxwell-Boltzmann distributions.** Energetic molecules of an ideal gas of mas *m*, freely move and collide in an isolated container (see Fig. 1). The Maxwell-Boltzmann distribution approximates well the molecules' macro-equilibrium (curve *T* on diagram). To rearrange such equilibrium motion with different speed distributions, say to separate the molecules in the container, into its two partitions, *L* with lower speeds (curve $T_L$ on diagram) and *H* with higher speeds (curve $T_H$ on diagram), it will require an effort with minimum due-work if achieved by the most efficient, reversible process (work-potential of such non-equilibrium). Note that partitioned molecules will always re-establish new distributions ($T_H$ and $T_L$ ), with extremely fast and extremely slow molecules in both partitions. There is no way (not even a 'demon') to create effortless non-equilibrium without "interference due-work" effort to suppress forced interference of other molecules (see also Fig. 1 & 3), in addition to and independently from the "gate-work" needed for demon's gate operation.

**Spontaneous molecular interactions have self-tendency towards thermal-equilibrium and against any otherwise process, like Maxwell's Demon (MD) selective gating.**

**Forceful, molecular interactions** would not "self-*pause to watch*", but **would spontaneously underline interfere** to "*compete and resist*" any other selective process, like **Maxwell's demon (MD) gating** a selected molecule **through a finite size hole** during a **finite gate-open time period**:

- **While one selected molecule is gated** through a hole into a partition, the other **thermal molecules (from both gate sides) would "compete" and spontaneously interfere** by forcefully resisting such a separation, so that **due-work would be required to prevent such interference** in order **for gating to be possible** (Fig. 1).

- The gated and partitioned molecules will tend, in each partition, to self-interact towards new equilibrium distributions, **always providing extremely-fast and extremely-slow molecules in both partitions** (Fig. 2).

- **If interference could be nonexistent**, an MD would, in long time, effortlessly separate molecules towards **infinity- and zero-temperature in the two partitions**, since there will always be extremely fast and extremely slow molecules in both partitions.

- However, **due to interference, MD cannot be used as an agent for perpetual work generation** (i.e., perpetual non-equilibrium generation) from within surrounding equilibrium, as **hypothetically alluded** by some, **in violation of established fundamental laws**.

- More **advanced, parametric MD simulations with realistic gate size and molecular speed, and accounting for thermal-interference**, may provide **answers to existing fallacies** and support phenomenological reasoning presented here.

- **Afterall**, **no experimental demon**, to perpetually generate non-equilibrium from within equilibrium, could have been realized **in over 150 years now**.

TEREFORE, relevant **major, interference "due-work"** (overlooked by Maxwell and his followers) **would be required to prevent interference** of other molecules with a gated molecule, **in addition to minor "gate-work"** to select a molecule at a time and to operate the gate.

**Fig. 3: Molecular natural-forcing or self-tendency towards thermal-equilibrium and "Interference Due-Work":** Molecular interactions spontaneously force towards thermal-equilibrium and against otherwise, that is they will interfere and obstruct (not "*self-pause to watch*") while any molecule is gated through a finite-size hole during a finite-time period. Relevant "*interference due-work*" will be required to prevent interference of other molecules, in addition to "*gate-work*" to just select a molecule and operate the gate. The former work is substantial and at least as the thermodynamic work-potential gained, while the latter work could be infinitesimally small if the gate operation is perfected to be virtually reversible.

While one selected molecule is gated through a hole into a partition, the other thermal molecules from both gate sides will spontaneously interfere by colliding, i.e., forcefully resisting such a separation and competing to pass through the gate while the gate is open. Therefore, a due-work would be required to prevent such interference in order for gating to be possible.

Furthermore, even if all molecules, faster than a chosen threshold value, are gated into one hotter-partition ($H$, with higher speeds) and equal number of slower molecules gated back into the other colder-partition ($L$, with lower speeds), they will spontaneously interact in each partition and rearrange into the respective natural-distributions, always with extremely high and extremely low speed values in both partition (curve $T$ 's on Fig. 2).

Another consequence of effortless MD operation would be that it will tend, given enough time, to separate molecules towards infinity- and zero-temperature in the two partitions, thus in limit resulting in all energy being available as work, since the partitioned molecules will tend to self-interact to a new equilibrium, near-Boltzmann's distributions, always providing extremely fast and extremely slow molecules in both partitions, see Fig. 1, 2 & 3. Effortless creation of such new miraculous-work-potential will be against natural forces and processes: it will be violation of the SL of thermodynamics, the latter is merely describing "natural-forcing or spontaneous self-tendency of nonequilibrium towards equilibrium", and impassibility otherwise. It would also have catastrophic consequences: the existence as we know it would not be possible.

A structure (like a boundary-contained body or any existing, self-sustained structure), in principle, created by external work, may perpetually retain non-equilibrium properties and work-potential in accord with the existing fundamental laws, but if restructured, it may provide only transient work limited by its work-potential (the minimum creation work in the most efficient reversible process), see Structural equilibrium section below. However, it cannot be used as an agent or agency for perpetual work generation (i.e., perpetual non-equilibrium generation) from within surrounding equilibrium, in violation of the existing fundamental laws, as hypothetically alluded by some.

Many MD's properties and interactions with all relevant gas molecules (i.e., thermal particles), container, partition and gate (i.e., a trapdoor) are conveniently oversimplified or ignored by demonologists. The most crucial fact, that the integral, chaotic interactions of all thermal particles on the MD's operation, has been overlooked, but focus on a single, opportunistic particle motion is emphasized, as if the other thermal particles would "*self-pause to watch*" and not interfere, see Fig. 1 & 3. Due effort to suppress such forced interference of other thermal particles would amount to required, major interference due-work, or '*due-work*' (further hereon for simplicity), to establish a macro non-equilibrium, which is independent and in addition to auxiliary '*gate-work*' of MD to observe molecules and operate a gate. The former, thermodynamic due-work, is unavoidable and substantial, while the latter, MD's operational work, could be infinitesimally small if the MD's operation is perfected to near-reversible actions, thus making illusion of the SL violation.

Furthermore, the hole, and therefore the gate, should be large enough to provide reasonable chance for any molecule to pass through (if it is about molecular size, the probability that any molecule will pass through will be virtually negligible). Moreover, the observation and gate actuation processes cannot be instantaneous to accommodate a single molecule only, but will require some finite time-period, large enough to achieve measurements and gate actuation (gate and actuator must be larger than molecular size), thus giving substantial chance to other chaotic and fast molecules to interfere and negate the MD's operation goal.

Therefore, for an MD, to succeed in separating faster from slower molecules in different partitions, it has to gate a molecule at a time, but also to prevent simultaneous interference of other thermal molecules, the latter overlooked by Maxwell and his followers. More advanced, parametric MD simulations with realistic gate size and molecular speed, and accounting for all interactions including unavoidable thermal interference, may provide answers to existing fallacies and support phenomenological reasoning presented here.

Moreover, since micro-properties and interactions are additive, after integration of MD's effects over time and space, from micro- to macro-system, for MD to somehow defy directions of natural, cause-and-effect, force-flux mass-energy displacement within macro-system, is nothing but wishful imagination.

No wonder that an operational Maxwell's demon could not have been successfully employed in last 150 years. It is reasoned here that an MD operation (a superficial being with imaginary and 'miraculous skills' and perfect tools) to effortlessly create a forced non-equilibrium from within an equilibrium, against natural forces and without due effort, is physically not possible.

## Original Maxwell's Demon and Its Fallacies

The original MD and its fallacy will be reasoned and discussed first. As conceived by Maxwell [1, 2], his miraculous being (MD) is schematically presented on Fig. 1. In Maxwell's words, "A [neat-fingered] 'being' whose attributes are essentially finite as our own, but is able to do what is at present impossible to us [thus miraculous], who can see the individual molecules, and open and close the hole [by operating a gate], so as to allow only the swifter [faster] molecules to pass from A to B [from $L$-low to $H$-high on Fig. 1, respectively and further hereon], and only the [same number of] slower ones to pass [back] from $H$ to $L$ [to maintain the same gas density]. He will thus, without expenditure of work, raise the temperature of $H$ and lower that of $L$ [thus original, thermal-MD], in contradiction to the Second law (SL) of thermodynamics." Maxwell was imagining a 'thought experiment' to somehow take advantage of the non-uniformity of speed of molecules of a gas in thermal equilibrium (Fig. 2), in order to 'effortlessly' separate them in two partitions, $L$ with lower molecular speed, and $H$ with higher molecular speed, and thus 'bypass' the ordinary forcing as observed in nature and stated by the SL of thermodynamics.

Furthermore, considering the requirements for the finite hole and gate size, considerably larger than a molecule size, and finite time-period for MD's operation, as well as the chaotic and fast molecular thermal motion (most molecules are faster than sound speed), it is probable that about the same speed molecule (also possible with higher or lower speeds) would pass back from $H$ to $L$ at that time, or collide with an oncoming molecule, see the dashed

"BUT!?" arrow line in $H$ part on Fig.1. Even higher speed molecules may and will pass back from $H$ to $L$, and more probably so if MD was 'successful by chance' to initially separate more high-speed molecules into $H$.

Therefore, just 'opening-and-closing' a gate, for a functional size hole and reasonable time period, would more 'equalize than separate' molecules by speed, similarly to providing any other 'opening' or removing the partition. As if Maxwell, focused on one opportunistic molecule, had 'overlooked to account for many other molecules', including from the other side molecules, that they will simultaneously compete to use the 'open-gate' opportunity. There would be the need for major "due-work" to prevent such forced interference by other chaotic molecules (i.e., thermal particles), obviously overlooked by Maxwell and his followers.

According to existing, well-established theory and practice, to create the thermal non-equilibrium with a gas in two partitions, one of lower temperature $T_L$ and the other of higher temperature $T_H$ than the original equilibrium temperature $T$, see Fig. 2, it will at minimum require a certain, finite work input, even with the most efficient, perfect processes, called here 'due-work'. That work, known as 'non-equilibrium work-potential', could be obtained back using ideal, reversible processes to bring back the achieved non-equilibrium to the original equilibrium.

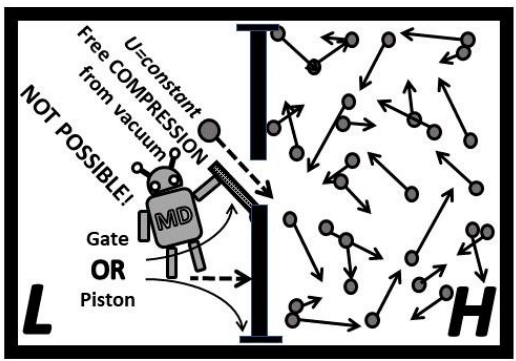

**Fig. 4: 'Free compression' is not possible.** An adiabatic gas system is presented with a 'density Maxwell's demon' (that allow passage of molecules to one side only). Such 'miraculous' MD would compress the expanded gas into the right-half partition-$H$ by opening the gate only when a gas molecule is approaching the gate to pass from the $L$-to-$H$ partition, spontaneously on their own (ideally without any external work). That would be equivalent to the piston compressing expanded gas into the partition-$H$, without any work. Such forceless 'free compression' would be against the gas pressure forces and thus physically impossible, since other gas molecules would be interfering and simultaneously passing both ways while the gate is open, see Fig.1 and 3.

Another 'thought experiment' with even simpler MD, which will only open the gate to allow molecules to pass from partition $L$ to $H$, but not in opposite direction (similar to one-way valve, also suggested by Smoluchowski [6]), is considered on Fig.4. In time, there will be increasingly more molecules in $H$ than in $L$, thus increasing density and pressure in $H$ until none of the molecules is left in $L$ (so-called, density-MD). The ideal gas energy (and temperature) will not change since no heat nor work was exchanged. This action will be equivalent to effortlessly compressing the molecules from $L$ to equal partition $H$, against the resulting pressure force in $H$, miraculously without any work-effort and isothermally, thus against the natural forces. Let us call this process (in adiabatic gas container) '*free compression*', which would obviously be super-natural and physically impossible. Such miraculous process would be 'ultra-reversible' where the gas work-potential would be miraculously generated and accompanied with entropy destruction while energy would be conserved. The compressed gas, generated by such MD without any work input, could reversibly expand back to initial pressure and extract its work-potential while subcooling to lower that the ambient temperature. Then such MD device may be spontaneously heated from the ambient and with repeated cyclic action perpetually generate work from ambient in thermal equilibrium. It would represent the "perpetual motion of the second kind (PMM2)," i.e., it would violate the Second law of thermodynamics.

Furthermore, the related reasoning and analyses by Szilard [7] and his followers, that MD would require the work for observation and information generation and storage, to avoid SL violation, is misplaced, since they were focused on MD's operation, to compensate related work and entropy reduction, but overlooking requirement for the due-work to prevent other molecules to interfere during the passage of the selected molecule, as detailed above. Not only that an MD cannot be effective without any work, it cannot be effective with an insufficient work, lower than the due-work. Moreover, the Szilard's one-molecule engine, an unrealistic imagination, cannot represent the multi-molecular chaotic, thermal phenomena, and therefore is physically meaningless. It violates SL even without any partition, as the single molecule moves and carry with it all properties from one partition to another, leaving 'vacuum' in previous one! Those are the fallacies of Szilard and his followers, that were recognized later, but unfortunately replaced with different kind of fallacies, see next.

**Local Destruction of Entropy Is Impossible and Cannot Be Compensated Elsewhere Later**

Landauer [8] and his followers, recognizing that the information and storage work suggested by Szilard and his followers is inadequate, introduced additional fallacies to save the SL. They stated that any MD's miraculous work-gain and related entropy reduction is compensated by follow-up memory information-erasure, accompanied with entropy generation. However, a fundamental law cannot be selectively violated and then 'saved' by compensation elsewhere and later, see Fig. 5. Entropy cannot be destroyed by MD, locally or at a time, and 'compensated' by generation elsewhere or later. It would be equivalent to allow rivers to spontaneously flow uphill and compensate it by more downhill flow later. We cannot pick-and-choose to violate a fundamental law and compensate it later elsewhere. Entropy is generated everywhere and always, at any scale without exception, and cannot be destroyed by any means at any scale [9, 10]. Impossibility of entropy reduction by destruction should not be confused with local entropy decrease due to entropy outflow with heat (thermodynamic entropy is associated with thermal motion or heat only, and caution should be exercised when it is defined at micro- and sub-micro scales). Entropy destruction would imply spontaneous increase of non-equilibrium and more efficient than a 100% reversible process, with mass-energy flux displacement against cause-and-effect, natural forces. It will also negate the reversible equivalency and existence of equilibrium.

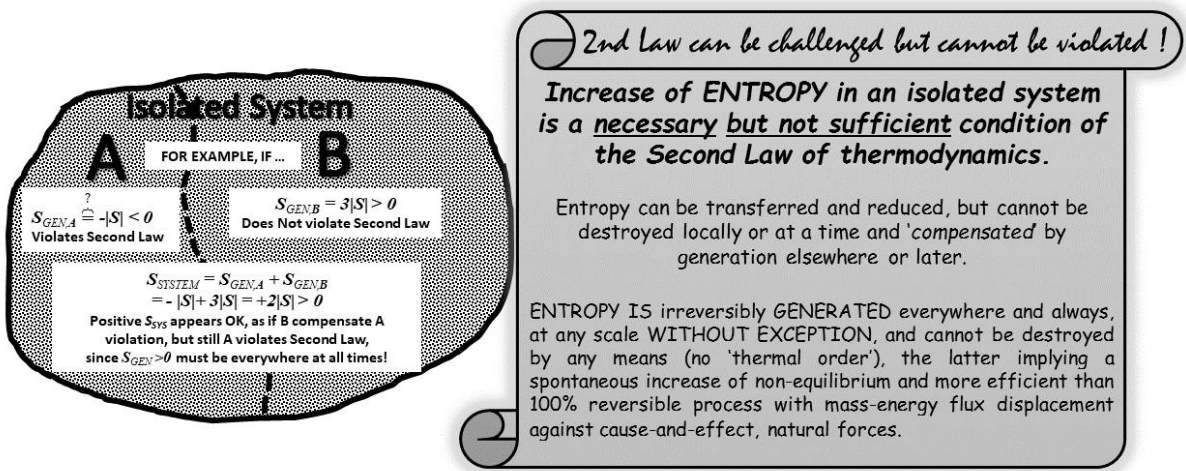

**Fig. 5: Destruction of entropy is impossible and cannot be 'compensated' elsewhere or at later time.** "Entropy of an isolated, closed system (or universe) is always increasing," is a necessary but not sufficient condition of the Second Law of thermodynamics. Entropy cannot be destroyed, locally or at a time, and then 'compensated' by generation elsewhere or later. It would be equivalent to allow a river to spontaneously flow uphill and compensate it by more downhill flow later. Entropy is generated everywhere and always, at any scale without exception, and cannot be destroyed by any means at any time or scale. Impossibility of entropy reduction by destruction should not be confused with local entropy decrease due to entropy outflow with heat.

Both, Szilard and Landauer (and their followers) misplaced their focus to justify MD's operational work *per se* if any, and any related entropy reduction, by so called 'compensation' by information generation and storage, and memory erasure accompanied by entropy generation, among others. However, they overlooked to justify thermodynamic due-work, generated freely (or below cost) by MD, as its main objective. After all, the MD is imagined 'to violate the SL', and all authors who try to exorcize MD by justifying its operation in accord with the SL, are in effect negate the possibility of MD and misuse its name. We should not confuse the ideal, 'thermodynamic due-work' (equal to the non-equilibrium work-potential) with any MD's 'operational gate-work' described above. That is the principal fallacy of all prior MD advocates, the authors who were trying to justify MD's miraculous possibility of violating the SL, by overlooking the need for the due-work, as well as fallacy of those authors who were trying to exorcize MD by unduly assigning a need for MD's operational, gate-work and equating it with thermodynamic due-work, to avoid the SL violation.

In ideal engines the entropy will be conserved and in real engines it will be irreversibly generated, regardless that thermal heat is reduced, i.e., converted to work. In summary, the ideal, reversible processes (thermal and/or non-thermal) will conserve entropy, and entropy would be generated only when work-potential energy, in part or in whole, is dissipated to thermal heat. When thermal energy is reduced by conversion to work (like in heat engine), entropy will not be reduced/destroyed since work is not related to entropy. Therefore, there is no way to destroy entropy [9, 10].

Furthermore, "Entropy of an isolated, closed system (or universe) is always increasing," is a necessary but not sufficient condition of the SL of thermodynamics, since entropy cannot be destroyed at any scale, locally and/or temporarily and compensated elsewhere and/or later.

**Homogeneous, Dynamic and Structural Equilibriums**

An isentropic, macro temperature oscillator, capable of producing perpetual temperature oscillations or perpetual temperature difference, is discussed and depicted on Fig. A1 in Appendix 1, thus demonstrating a thermal-mechanical, "structural equilibrium" with non-uniform temperature, or "dynamic equilibrium" with perpetual temperature oscillations, without violating the SL. Therefore, the fluctuation phenomena do not violate the SL since the temperature fluctuations could be adiabatic (in limit, isentropic) or due to heat transfer fluctuations. Therefore, such quasi-equilibrium with non-uniform temperature cannot be utilized for perpetual work generation without perpetual work consumption from elsewhere. Similarly, the perpetual dynamic oscillations or fluctuations at micro or macro scales may not be utilized for perpetual work generation. Therefore, only limited transient, but not a perpetual work, may be obtained from such oscillating or fluctuating systems.

To summarize, there are three types of self-sustained macro equilibriums (stable and perpetual, with balanced macro-forces, including inertial forces, and net-zero mass-energy fluxes): (1.) *homogeneous equilibrium* with uniform-properties, e.g., thermo-mechanical equilibrium within simple fluids and solids, as in classical thermodynamics; (2.) *dynamic equilibrium* (a.k.a. quasi-equilibrium) with spontaneous and perpetual self-sustained motion, like ideal thermal oscillations (as on Fig. A1 in Appendix), ideal pendulum oscillations, orbiting electrons around a nucleus, thermal-molecular micro random-motion, etc.; and (3.) *structural equilibrium* (a.k.a. "boundary constrained') with heterogeneous and non-uniform properties, like mechanical, thermal, electrical and chemical potentials, i.e.: (i) hydrostatic pressure in gravity field, (ii) inflated air-balloon; (iii) hot medium in an adiabatic container (e.g., thermos flask); (iv) electro-chemical cell; (v) fuel-air mixture (ready for combustion), etc.

Dynamic and structural equilibriums with non-uniform properties, as compared with classical 'homogeneous equilibrium', are elusive and may be construed as non-equilibriums, if they are re-structured and in a transient process some useful work is obtained, to allude to violation of the SL. However, such work-potential is limited to one stored within such a structure (regardless how small or large), and it cannot be utilized as a perpetual (stationary or cyclic) PMM2 (perpetual motion machine of the second kind) device or MD, to continuously generate useful work from within an equilibrium.

Most of the fallacies of diverse MDs and SL violations are related to elusive structural quasi-equilibriums when transient work potentials are hypothesized to be perpetual. Engaging any engine or engine-like structure to produce work may run only transiently until existing physical and electro-chemical work-potentials are exhausted and another equilibrium is re-established with such a new device or structure. Testing such hypothetical devices, to produce perpetual work from within surrounding equilibrium, would be straight-forward and not difficult, but somehow overlooked by the SL challengers. However, a PMM2-SL violation has never been confirmed possible, but to the contrary. Therefore, speculating that such devices (or engines, or MDs) will work perpetually and violate the SL is only an illusory imagination (a "thought experiment") and physically impossible.

## Maxwell's Demon Infinitely-efficient Refrigerator and Its Catastrophic Consequences

If Maxwell' demons could produce a perpetual thermal non-equilibrium within a system in thermal equilibrium (e.g., spontaneous cooling within the surroundings), by separation of faster from slower molecules, taking advantage of such molecular speed distribution within equilibrium, then such a refrigerator with MDs would have infinite cooling efficiency (i.e., would produce perpetual cooling without any work input, *Cooling efficiency* $= Q_{COOLING}/W_{MD}$ $= Q_{COOLING}/0 = infinity$).

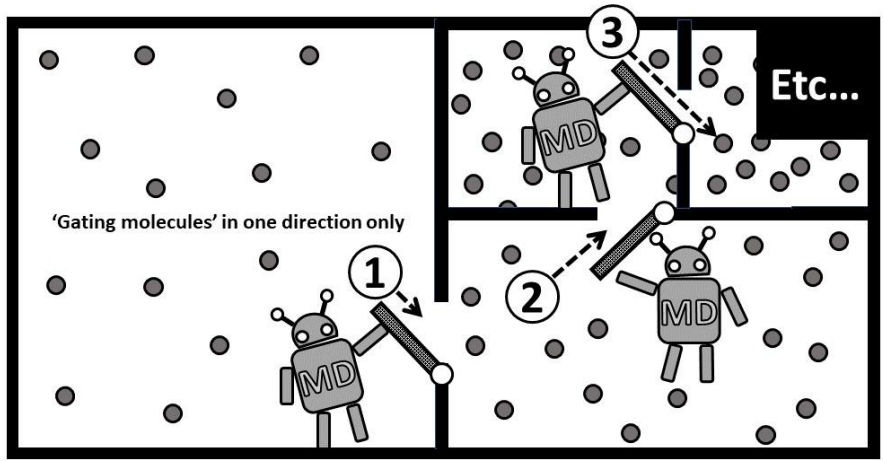

**Fig. 6:** '**Would-be a catastrophic consequence' if Maxwell's demon is possible.** 'Density Maxwell's demons' (that allow passage of molecules to one side only, like one-way valve), will in sequence "1-2-3-Etc…" concentrate density and energy into an infinitesimal-corner volume with infinite concentration, a catastrophic consequence of smart-and-effortless, but luckily impossible demons.

By spontaneously creating thermal non-equilibrium from within an equilibrium the MDs will in fact negate the very existence of such thermal equilibrium. If such MD refrigerator is used as a perpetual heat sink for a heat engine, it will also be infinitely efficient, i.e., it could produce perpetual, free work by using heat from the surroundings, in violation of the SL of thermodynamics.

Not only that MD will violate the SL by producing free useful work, but it could produce some unwanted or even catastrophic consequences. On Fig. 6, a '*Would-be a catastrophic consequence*' with MDs is depicted. So-called 'density MDs', that allow passage of molecules to one side only (like one-way valve; it would also be much simpler to operate than the original, 'thermal MD'), will in sequence "1-2-3-Etc…" concentrate density and energy into an infinitesimal-corner volume with infinite density and energy concentration, a catastrophic consequence of smart-and-effortless, but luckily impossible demons.

## Discussions and Conclusion

With phenomenological insight and simple reasoning, the Maxwell's demon (MD) is demystified by accounting for overlooked "interference due-work", see summary on Fig. 7. Namely, an MD cannot effortlessly operate the gate to pass a molecule at a time, due to simultaneous interference of all other molecules which will "compete (collide and pass through) to use the open gate opportunity," and thus more equalize than generate thermal non-equilibrium, since the random thermal motion does not have any directional preference for gating to work alone. However, the effort to prevent simultaneous interference of all other molecules (i.e., the necessary, major "interference due-work"), to enable MD gating, was overlooked by Maxwell and his followers, as well as the MD exorcists, who tried to inappropriately justify the MD operation. Further research, including more comprehensive experimental verification and more realistic molecular dynamic and statistical simulations, should attest to the above phenomenological reasoning.

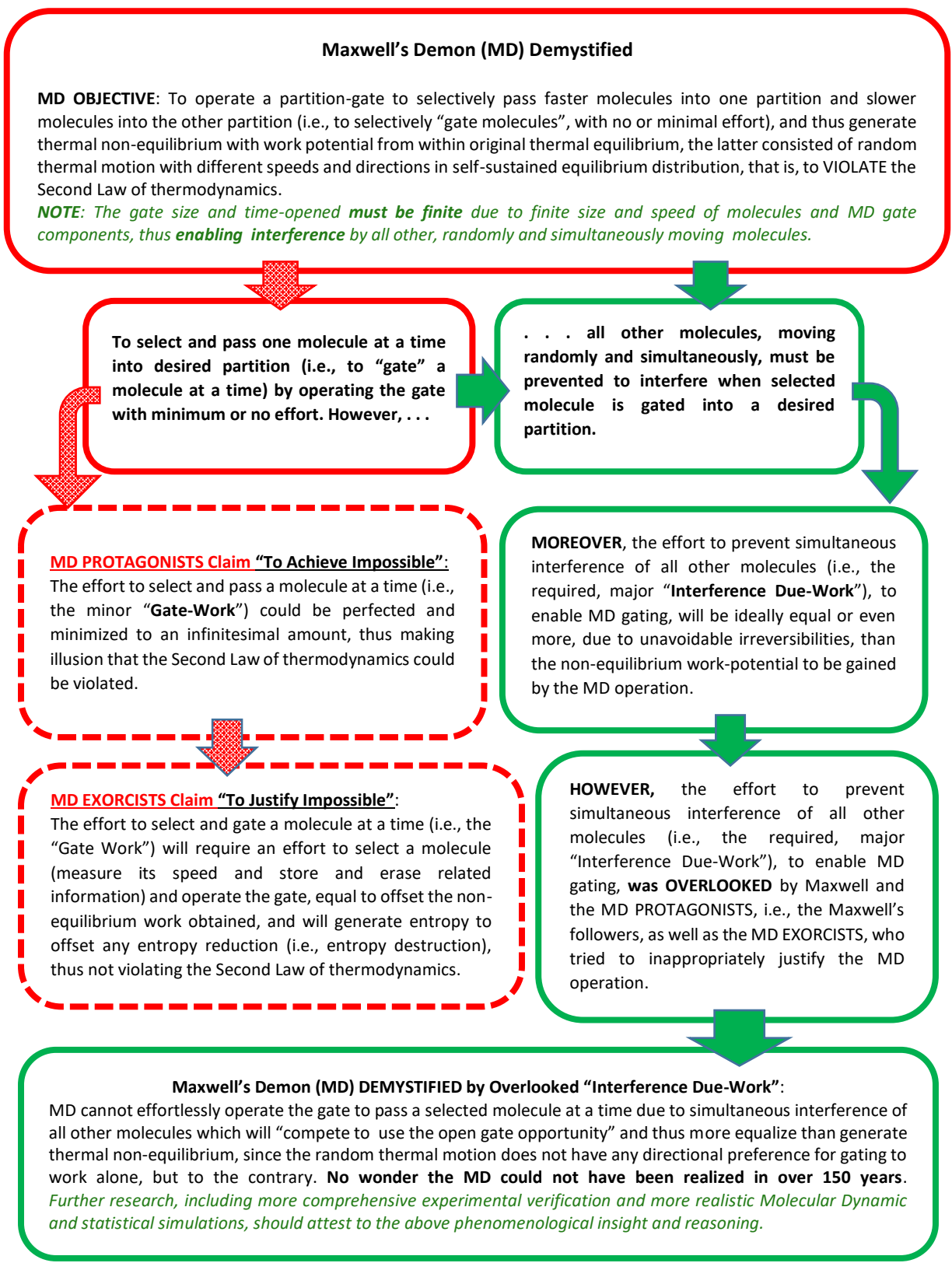

**Maxwell's Demon (MD) Demystified**

**MD OBJECTIVE**: To operate a partition-gate to selectively pass faster molecules into one partition and slower molecules into the other partition (i.e., to selectively "gate molecules", with no or minimal effort), and thus generate thermal non-equilibrium with work potential from within original thermal equilibrium, the latter consisted of random thermal motion with different speeds and directions in self-sustained equilibrium distribution, that is, to VIOLATE the Second Law of thermodynamics.
*NOTE: The gate size and time-opened **must be finite** due to finite size and speed of molecules and MD gate components, thus **enabling interference** by all other, randomly and simultaneously moving molecules.*

To select and pass one molecule at a time into desired partition (i.e., to "gate" a molecule at a time) by operating the gate with minimum or no effort. However, . . .

. . . all other molecules, moving randomly and simultaneously, must be prevented to interfere when selected molecule is gated into a desired partition.

**MD PROTAGONISTS Claim "To Achieve Impossible"**:
The effort to select and pass a molecule at a time (i.e., the minor "Gate-Work") could be perfected and minimized to an infinitesimal amount, thus making illusion that the Second Law of thermodynamics could be violated.

**MOREOVER**, the effort to prevent simultaneous interference of all other molecules (i.e., the required, major "**Interference Due-Work**"), to enable MD gating, will be ideally equal or even more, due to unavoidable irreversibilities, than the non-equilibrium work-potential to be gained by the MD operation.

**MD EXORCISTS Claim "To Justify Impossible"**:
The effort to select and gate a molecule at a time (i.e., the "Gate Work") will require an effort to select a molecule (measure its speed and store and erase related information) and operate the gate, equal to offset the non-equilibrium work obtained, and will generate entropy to offset any entropy reduction (i.e., entropy destruction), thus not violating the Second Law of thermodynamics.

**HOWEVER**, the effort to prevent simultaneous interference of all other molecules (i.e., the required, major "Interference Due-Work"), to enable MD gating, **was OVERLOOKED** by Maxwell and the MD PROTAGONISTS, i.e., the Maxwell's followers, as well as the MD EXORCISTS, who tried to inappropriately justify the MD operation.

**Maxwell's Demon (MD) DEMYSTIFIED by Overlooked "Interference Due-Work"**:
MD cannot effortlessly operate the gate to pass a selected molecule at a time due to simultaneous interference of all other molecules which will "compete to use the open gate opportunity" and thus more equalize than generate thermal non-equilibrium, since the random thermal motion does not have any directional preference for gating to work alone, but to the contrary. **No wonder the MD could not have been realized in over 150 years**.
*Further research, including more comprehensive experimental verification and more realistic Molecular Dynamic and statistical simulations, should attest to the above phenomenological insight and reasoning.*

**Fig. 7**: Maxwell's Demon (MD) Demystified * ©2019 by M. Kostic * [www.kostic.niu.edu](www.kostic.niu.edu)

Many MD-like devices in the literature have been product of creative imagination (i.e., thought experiments), always with idealization and limitation of system properties and processes, and without complete analysis or proper experimental verification. No wonder that a functional demon, with real perpetual utility, could not have been achieved in 150 years since its inception. Critical, underlying mass-energy structures and processes at micro- and sub-micro scales are more complex and sometimes undetected at our present state of tooling and mental comprehension. However, their integral manifestations at macroscopic level are more realistically observable and reliable, thus being the check-and-balance of statistical microscopic and quantum hypotheses.

Reversible fluctuations or oscillations at sub-micro, micro- or macro scales are not violation of the SL. Actually, entropy could fluctuate (reduce-and-increase locally and temporally) due to entropy transfer (with heat transfer) within interacting particles and/or systems. Reversibility is the 'true' equivalency (nothing is lost nor degraded), where momentum and energy transfers are the most efficient, and any expended work could be entirely recovered with reversed, thus most efficient reversible processes, rendering all reversible processes to be the most, 100%, and equally efficient. However, any super-efficiency, including MD's, above maximum 100% would negate the reversible equivalency. The well-established and universal, thermodynamic reversible-equivalency cannot be ignored at any scale nor with any device. Otherwise any special violation will imply the general violation of the SL. Molecular speed-separation by MD's timing is simply against the physics of the thermal motion, since there is no consistent directional preference for timing to be effective and achieve the separation goal.

The fundamental physical laws are independent from any system structure or scale, and they should take primacy over any special analysis based on approximations and limitations of modeling of system, its properties and processes, and especially if based on 'thought experiments'. After all, the micro- and sub-micro simulations and experimental analyses are also based on the fundamental Laws and therefore cannot be used to negate those fundamental Laws. As the fundamental laws of nature and thermodynamics are expanded from simple systems in physics and chemistry, to different space and time scales and to much more complex systems in biology, life and intelligent processes, there are more challenges to be comprehended and understood.

## REFERENCES

1.  C.G. Knott, Life and Scientific Work of Peter Guthrie Tait (Cambridge University Press, London, pp. 213-215 (1911): Maxwell had introduced this idea in a 1867 letter to Peter Guthrie Tait "… to pick a hole" in the Second law.

2.  J.C. Maxwell, Theory of Heat; Longmans, Green, and Co.: London, UK, 1871; Chapter 12, (1871).

3.  W. Tompson, "The Kinetic Theory of the Dissipation of Energy," Nature 9, 441-444 (1874).

4.  J. Maddox, Maxwell's demon Flourishes, Nature; 345, 6271 (1990).

5.  H. S. Leff and A.F. Rex (eds.), Maxwell's demon 2: Entropy, Classical and Quantum Information, Computing. CRC Press. ISBN 0-7503-0759-5 (2002).

6.  M. Smoluchowski, Experimentell nachweisbare der üblichen Thermodynamik wiedersprechende Molekularphänomene. Physikalische Zeitschrift. 13, 1069–1080 (1912, in German).

7.  L. Szilard, Über die Entropieverminderung in einem thermodynamischen System bei Eingriffen intelligenter Wesen (On the reduction of entropy in a thermodynamic system by the intervention of intelligent beings)". Zeitschrift für Physik. 53: 840–856, (1929, in German) [English translation available as NASA document TT F-16723 published 1976].

8.  R. Lindauer, Irreversibility and heat generation in the computing process. IBM J. Res. Dev., 5, 183–191 (1961).

9.  M. Kostic, Revisiting the Second Law of Energy Degradation and Entropy Generation: From Sadi Carnot's Ingenious Reasoning to Holistic Generalization. *AIP Conf. Proc*. 1411, 327 (2011). DOI:10.1063/1.3665247.

10. M. Kostic, The Elusive Nature of Entropy and Its Physical Meaning, Entropy, 16(2) 953-967 (2014); DOI:10.3390/e16020953. http://www.mdpi.com/1099-4300/16/2/953/ (accessed January 2020).

## APPENDIX 1: Dynamic and Structural Equilibriums

An isentropic, macro temperature oscillator, capable of producing perpetual temperature oscillations or perpetual temperature difference, is depicted on Fig. A1. If a piston, in the cylinder with ideal, inertial mass and elastic spring system, at isothermal center-position, is compressing an ideal gas in partition B, thus isentropically increasing the gas temperature, then gas will expand in partition A and isentropically decrease its temperature. If the displaced piston is left free, it will 'perpetually oscillate', but without any perpetual work generation, similarly to an ideal pendulum oscillation, thus demonstrating a thermal 'dynamic equilibrium' with perpetual temperature oscillations. At any locked, stationary piston position, a self-sustained 'structural equilibrium' will establish with perpetual temperature difference, as depicted on Fig. A1, without violating the SL. The initial compression work may be obtained back if original equilibrium is re-established in a transient reversible process. However, such non-uniform temperature cannot be utilized for perpetual work generation without perpetual work consumption from elsewhere.

Similarly, the perpetual dynamic oscillations or fluctuations at micro or macro scales may not be utilized for perpetual work generation. It is shown on Fig. A1 that even macro fluctuations, after a transient start-up process, may be set into perpetual oscillatory motion without any external perpetual work input or output. Therefore, only limited transient, but not a perpetual work, may be obtained from such oscillating or fluctuating systems. As depicted on Fig. A1, an ideal adiabatic piston-cylinder with gas, may demonstrate a thermal-mechanical, structural equilibrium with non-uniform temperature or perpetual temperature oscillations, without violating the SL.

Fluctuating phenomena in perfect equilibrium are reversible, thus isentropic, so that any reduction in entropy reported in the literature (whatever that means), may be due to 'improvised' and incomplete entropy definitions at

micro- and sub-micro scales, approximate accounting for thermal and/or neglecting diverse displacement contributions to entropy (for example, during isothermal free expansion of an ideal gas, entropy is generated due to elastic-field volume-displacement, $\delta S_{gen}=dS=R_{gas}dV/V$, and similarly could be due to subtle and illusive micro- and sub-micro forced-field 'displacements', including particle 'correlation' and/or quantum entanglement in respective force fields, etc.). Also, note that fluctuating temperature does not always mean fluctuation of entropy as demonstrated by the isentropic compression-expansion on Fig. A1, etc. Therefore, the fluctuation phenomena do not violate the SL since the temperature fluctuations could be adiabatic (in limit, isentropic) or due to heat transfer fluctuations.

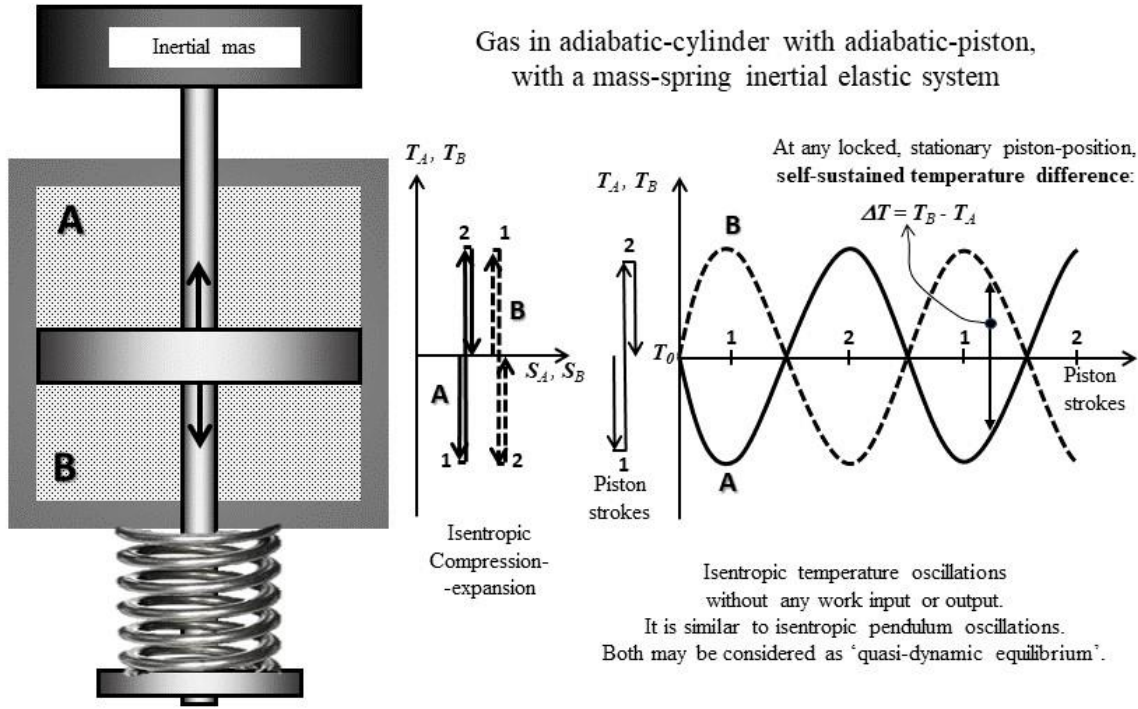

**Fig. A1: Gas in adiabatic piston-cylinder system isentropically produces perpetual temperature oscillations or temperature difference.** If the depicted piston, from isothermal center-position, is compressing an ideal gas in partition B, thus isentropically increasing the gas temperature, then gas will expand in partition A and isentropically decrease its temperature, without any heat transfer. If displaced piston within the cylinder with ideal, inertial mass and elastic spring system is left free, it will perpetually oscillate, but without any perpetual work generation, similarly to an ideal pendulum oscillations, thus demonstrating a thermal 'dynamic equilibrium' with perpetual temperature oscillations (but not entropy oscillations); or, at any locked, stationary piston position, a self-sustained 'structural equilibrium' will establish with perpetual temperature difference, without violating the Second law.

In summary, there are three types of self-sustained macro equilibriums (stable and perpetual, with balanced macro-forces, including inertial forces, and net-zero mass-energy fluxes): (1) *homogeneous equilibrium* with uniform-properties, e.g., thermo-mechanical equilibrium within simple fluids and solids, as in classical thermodynamics; (2) *dynamic equilibrium* (a.k.a. quasi-equilibrium) with spontaneous and perpetual self-sustained motion, like ideal thermal oscillations on Fig. A1, ideal pendulum oscillations, orbiting electrons around a nucleus, thermal-molecular micro random-motion, etc.; and (3) *structural equilibrium* (a.k.a. "boundary constrained') with heterogeneous and non-uniform properties, like mechanical, thermal, electrical and chemical potentials, i.e.: (i) hydrostatic pressure in gravity field, (ii) inflated air-mattress; (iii) hot medium in an adiabatic thermos flask; (iv) electro-chemical cell; (v) fuel with air (ready for combustion), etc.

Dynamic and structural equilibriums with non-uniform properties, as compared with classical 'homogeneous equilibrium', are elusive and may be construed as non-equilibriums, if they are re-structured and in a transient process some useful work is obtained, to allude to violation of the SL. However, such work-potential is limited to one stored within such a structure (regardless how small or large), and cannot be utilized as a perpetual (stationary or cyclic) PMM2 device or MD, to continuously generate useful work from within an equilibrium.